\documentclass[preprint,showpacs,preprintnumbers,amsmath,amssymb]{revtex4}
\usepackage[dvips]{graphicx}
\usepackage{amsmath}
\usepackage{bm}

\begin{document}

\preprint{BA-02-38}

\title{Atmospheric neutrino mixing and $b$-$\tau$ Unification}

\author{S.M. Barr}
\email{smbarr@bxclu.bartol.udel.edu}
\author{I. Dorsner}
\email{dorsner@physics.udel.edu}
\affiliation{
Bartol Research Institute\\
University of Delaware\\
Newark, DE 19716}

\begin{abstract}
Extrapolating the $\tau$ and $b$ masses in the MSSM tends to give for 
their ratio at the GUT scale a number around 1.2, for most viable values 
$\tan \beta$, rather than the minimal $SU(5)$ prediction of 1. 
We suggest that this may be due to large off-diagonal elements in the
charged lepton mass matrix $M_L$ that can also explain the large atmospheric
neutrino mixing angle. Several simple models with definite predictions 
for $m_{\tau}/m_b(M_{GUT})$ are presented.
\end{abstract}


\maketitle

\newpage
In minimal $SU(5)$ and many
other simple schemes of grand unification it is predicted that
$m_b = m_{\tau}$ at the unification scale \cite{btau1,btau2}. This relation 
is known to work rather well in the context of the MSSM \cite{susy1,susy2}. 
That is, when 
the actual masses of the $b$ and $\tau$ measured at low energy are
extrapolated in the MSSM to $M_{GUT}$, they are found to be nearly equal. 
Nevertheless, for a wide range of $\tan \beta$, $m_{\tau}/m_b(M_{GUT})$ 
tends to be about 1.2 rather than 1.0, as is shown in Table~\ref{tab:table1}. In this paper
we suggest that $m_{\tau}$ may be slightly larger than $m_b$ at the
GUT scale because of the large mixing in the lepton sector between the
third family and the lighter families that is seen in atmospheric
neutrinos. 

First, let us review briefly the situation with regard to $b$-$\tau$ 
unification in the MSSM. While it is true that for most of the range of 
$\tan \beta$ the ratio $m_{\tau}/m_b(M_{GUT})$ deviates substantially 
from unity, it is well known that this deviation is small if $\tan \beta$ 
is either close to $m_t/m_b$ or close to 1 (the so-called high $\tan \beta$
and low $\tan \beta$ allowed regions) \cite{tanbeta1,tanbeta2,tanbeta3}. 
The reason these regions give better
agreement with the minimal $SU(5)$ prediction is that they correspond to
certain Yukawa couplings becoming large enough to reduce the value of
$m_b$ slightly through their effect on its renormalization group running.
(For $\tan \beta \cong m_t/m_b$, the bottom Yukawa coupling is nearly equal to 
that of top, and therefore of order one; and for 
$\tan \beta \cong 1$, the top Yukawa coupling becomes greater than one
due to the fact that $v_u = v (1 + (\tan \beta)^{-2})^{-1/2}$ 
drops significantly below $v$.) 

The high $\tan \beta$ case is attractive from the viewpoint of minimal
$SO(10)$ unification, where the two MSSM Higgs doublets come from a
single $SO(10)$ multiplet, implying that the top and bottom Yukawa 
couplings are equal at the GUT scale \cite{als}. However, in more realistic 
$SO(10)$ models that attempt to fit the quark and 
lepton mass spectrum it is often not the case that the two MSSM doublets  
come purely from a single $SO(10)$ multiplet, and consequently 
$\tan \beta$ becomes a free parameter \cite{abb,bpw}. Moreover, from 
the point of view of the MSSM, a value of $\tan \beta$ as large as 
$m_t/m_b$ involves somewhat of a 
fine-tuning \cite{high}. As far as the low $\tan \beta$ case is concerned, 
values of $\tan \beta$ very close to 1 are now excluded in the constrained
MSSM \cite{low1,low2}.

Another way to save the minimal $SU(5)$ prediction, besides assuming
extreme values of $\tan \beta$, is to invoke finite radiative
corrections to $m_b$ coming from gluino and chargino loops
(the gluino loop being typically the larger) \cite{oneloop1,oneloop2}. 
Since there are contributions to these loops that are proportional 
to $\tan \beta$, they can be substantial. For $\tan \beta \approx 30$,
for instance, $m_b$ can easily receive a net correction of 15\% to 20\%,
which, assuming it is negative, would restore agreement with minimal $SU(5)$. 
However, this possibility is not without difficulties. The sign of the 
gluino and chargino loops are given, respectively, by $sgn(\mu M_3)$ and 
$-sgn(\mu A_t)$. Generally, to lower the value of $m_b$ and improve 
agreement with the minimal $SU(5)$ prediction, one needs that $sgn(\mu M_3)$
be negative, which would typically imply in minimal supergravity that
$sgn(\mu A_t)$ also is negative. However,
if $sgn(\mu A_t)$ is negative, then the chargino-loop contributions to the 
$b \rightarrow s \gamma$ amplitude adds constructively to the 
charged-Higgs-loop and standard model contributions, giving too large an effect
unless sparticle masses are assumed to be large. (If $sgn(\mu A_t)$ is 
positive, on the other hand, then the charged-Higgs-loop and the 
chargino-loop contributions to $b \rightarrow s \gamma$ tend to cancel
each other, which is good since the experimental result for 
$b \rightarrow s \gamma$ is consistent with the SM prediction.) 
For a recent discussion
of the one-loop SUSY corrections to $m_b$ and the constraints on them coming
from $b \rightarrow s \gamma$, see \cite{ky}.

Thus, although there are ways to save the minimal $SU(5)$ prediction
for $m_{\tau}/m_b$, perhaps the most conservative assumption is that
this ratio is indeed slightly larger than unity at the GUT scale.  
In this paper we explore one reason why this might be so.
We shall propose several versions of this idea that make definite predictions
for $m_{\tau}/m_b(M_{GUT})$. Eventually, if supersymmetry is discovered,
it may be possible to determine the values of $\tan \beta$ and the parameters
that are needed to compute the gluino- and chargino-loop corrections to
$m_b$, and thus pin down the value of $m_{\tau}/m_b(M_{GUT})$ and test 
these models.

The basic idea of this paper is that $m_{\tau}$ is made slightly
larger than $m_b$ due to the same mixing effects in the lepton sector
that produce the large neutrino mixing angle $\theta_{atm}$.
(Indeed, in some published models that explain the large atmospheric neutrino
mixing, it does happen that $m_{\tau}$ is slightly larger than $m_b$ at the
GUT scale. For instance, by a factor of 1.04 in \cite{abb} and a factor
1.08 in \cite{bpw}. These papers to some extent inspired the present
work.)

Suppose, for example, that the charged-lepton and down-quark mass matrices
have these approximate forms at the GUT scale:
\begin{equation}
M_L \cong \left( \begin{array}{ccc} 0 & 0 & 0 \\ 0 & 0 & \rho \\
0 & 0 & 1 \end{array} \right) \; m, \;\;\;\; 
M_D \cong \left( \begin{array}{ccc} 0 & 0 & 0 \\ 0 & 0 & 0 \\
0 & 0 & 1 \end{array} \right) \; m,
\end{equation}
\noindent
where $\rho \sim 1$.
The zeros represent small elements that give the masses of the lighter
two families. These matrices imply that $m_{\tau}/m_b \cong \sqrt{1 + \rho^2}$. 
The large off-diagonal element $\rho$ in $M_L$ also produces a large
mixing angle between $\tau$ and $\mu$, namely $\tan \theta^{\ell}_{23} = \rho$.
If we assume that the neutrino mass matrix is nearly diagonal
or hierarchical in this basis, so that the leptonic mixing angles come almost
entirely from the charged lepton sector, then
$\tan \theta_{atm} \cong \tan \theta^{\ell}_{23}$, implying that
\begin{equation}
m_{\tau}/m_b(M_{GUT}) \cong \sqrt{1 + \tan^2 \theta_{atm}} \cong \sqrt{2}.
\end{equation}

This kind of ``lopsided" form for $M_L$ has been suggested by many groups
as a simple way to explain the largeness of the atmospheric neutrino
mixing angle \cite{lopsided1,lopsided2,lopsided3,lopsided4}. 
A ``bimaximal" (or perhaps it is better to say 
``bi-large") pattern of neutrino mixings can be elegantly explained by a simple
extension of this idea \cite{bimaximal}, namely assuming that a whole 
column of $M_L$ is large:
\begin{equation}
M_L \cong \left( \begin{array}{ccc} 0 & 0 & \rho' \\ 0 & 0 & \rho \\
0 & 0 & 1 \end{array} \right) \; m,
\end{equation}
\noindent
where $\rho' \sim \rho \sim 1$.
Under the same assumption about the neutrino mass matrix, this gives
$\tan \theta_{sol} \cong \tan \theta^{\ell}_{12} = \rho'/\rho \sim 1$,
and $\tan \theta_{atm} \cong \tan \theta^{\ell}_{23} = 
\sqrt{\rho^2 + \rho^{\prime 2}} \sim 1$, with the 13 leptonic mixing angle
$U_{e3}$ small. If $M_D$ has the form in Eq. (1), then 
$m_{\tau}/m_b \cong \sqrt{1 + \rho^2 + \rho^{\prime 2}}$,
giving again the relation shown in Eq. (2). 

The value $\sqrt{2}$ for $m_{\tau}/m_b(M_{GUT})$ may seem too high, but 
if the one-loop SUSY correction to $m_b$ is around 
15\% or 20\%, with the {\it positive} sign suggested by 
$b \rightarrow s \gamma$, then it is quite consistent with the experimental
mass ratio for a wide range of $\tan \beta$

The matrices given in Eq. (1) and (3) give $m_{\tau}$ larger than $m_b$ at the
GUT scale because of the difference in form between $M_L$ and $M_D$.
However, it is well-known that in minimal $SU(5)$ these matrices are
transposes of each other. If $M_D \neq M_L^T$, then some non-minimal 
Yukawa structure must be involved. There are several simple possibilities.
One is that the off-diagonal elements $\rho$ and, in the case of Eq. (3), 
$\rho'$ come from a $\overline{{\bf 45}}$ of Higgs. This introduces a
relative factor of -3 between $M_L$ and $M_D$ elements as pointed out 
by Georgi and Jarlskog long ago \cite{gj}. 
The matrices of Eq. (1) would then become
\begin{equation}
M_L \cong \left( \begin{array}{ccc} 0 & 0 & 0 \\ 0 & 0 & \rho \\
0 & 0 & 1 \end{array} \right) \; m, \;\;\;\; 
M_D \cong \left( \begin{array}{ccc} 0 & 0 & 0 \\ 0 & 0 & 0 \\
0 & -\frac{1}{3} \rho & 1 \end{array} \right) \; m,
\end{equation}
\noindent
Making the same assumption that the atmospheric mixing is almost entirely
coming from $M_L$ (i.e. in this basis $M_{\nu}$ is nearly diagonal
or hierarchical), we then have 
\begin{equation}
m_{\tau}/m_b(M_{GUT}) \cong 
\frac{\sqrt{1 + \tan^2 \theta_{atm}}}{\sqrt{1 + \tan^2 \theta_{atm}/9}} 
\cong 3/\sqrt{5} \cong 1.34.
\end{equation}
\noindent
The same prediction results if this is generalized to bi-maximal
mixing.

\begin{figure}
\begin{center}
\includegraphics[width=4in]{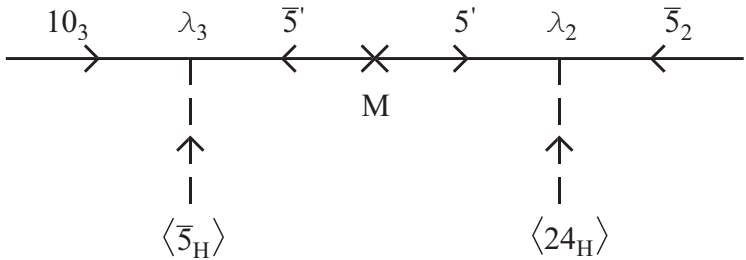}
\end{center}
\caption{\label{brane1} A diagram that can give operators producing
``lopsided" contributions to $M_D$ and $M_L$ in the usual four-dimensional setting.
The cross represents the explicit mass term of the vectorlike pair $\overline{{\bf 5}}' + {\bf 5}'$.}
\end{figure}

A way that a different Clebsch might arise is through an effective 
operator involving an adjoint Higgs (${\bf 24}$) field. Specifically, 
suppose that there is 
a superheavy vectorlike pair of quark and lepton multiplets, 
denoted $\overline{{\bf 5}}' + {\bf 5}'$, and that the superpotential
contains the following couplings:
\begin{equation}
W \supset \lambda_{33} {\bf 10}_3 \overline{{\bf 5}}_3 \overline{{\bf 5}}_H
+ M \overline{{\bf 5}}' {\bf 5}' 
+ \lambda_2 \overline{{\bf 5}}_2 {\bf 5}' {\bf 24}_H
+ \lambda_3 {\bf 10}_3 \overline{{\bf 5}}' \overline{{\bf 5}}_H 
\end{equation}
\noindent
Integrating out the heavy vectorlike fields, as shown in Fig.~\ref{brane1},
the resulting effective operator has the form, for 
$\langle {\bf 24}_H \rangle/M$ small, ${\bf 10}_3 (\overline{{\bf 5}}_2
{\bf 24}_H) \overline{{\bf 5}}_H/M$ (where the fields inside the parentheses
are contracted into a $\overline{{\bf 5}}$). This leads to the off-diagonal
element we have called $\rho$, but with a relative Clebsch of $-\frac{2}{3}$ 
between $M_D$ and $M_L$ (i.e. replace the $-\frac{1}{3}$ in Eq. (4) by
$-\frac{2}{3}$). This gives
\begin{equation}
m_{\tau}/m_b(M_{GUT}) \cong 
\frac{\sqrt{1 + \tan^2 \theta_{atm}}}{\sqrt{1 + 4 \tan^2 \theta_{atm}/9}} 
\cong \sqrt{18/13} \cong 1.17.
\end{equation}

We have been discussing $b$-$\tau$ unification in the context 
of SUSY $SU(5)$ grand unification. However, as is well known, four-dimensional
SUSY $SU(5)$ models typically have difficulties with natural doublet-triplet
splitting and proton decay from dimension-5 operators \cite{pdk}. 
(The former can be resolved by means of the missing partner mechanism, by 
introducing Higgs in 50 and 75 dimensional representations \cite{missing1,missing2}.) 
As recent papers have shown,
these problems can be simply evaded in SUSY GUT models with one or two
extra space dimensions compactified on orbifolds, the orbifold fixed points
being three-branes \cite{orbifold1,orbifold2}. 
Very simple and realistic SUSY $SU(5)$ models can be 
constructed with only a single extra dimension, compactified on
$S^1/Z_2 \times Z_2$. Recently, Hall and Nomura have pointed out that
extending SUSY $SU(5)$ to five-dimensions in the simplest way can  
improve the agreement with the experimental value of $\alpha_s$ \cite{hn}. 
In light 
of these facts, it seems that a more satisfactory context in which to
discuss $b$-$\tau$ unification may be five-dimensional models. 

We will now discuss a simple five-dimensional model that implements 
the idea contained in Eq. (1). This model is very similar to the 
four-dimensional model of Eq. (6). As in that model, there is an 
extra vectorlike $\overline{{\bf 5}} + {\bf 5}$ pair of quark/lepton 
multiplets that, when integrated out, leads to the off-diagonal element 
$\rho$. However, in the five-dimensional model this vectorlike pair does 
not couple to an adjoint (${\bf 24}$) of Higgs, as in Eq. (6), 
but feels the breaking of $SU(5)$ simply due to the fact that it lives in 
the five-dimensional bulk.

Imagine an $N=1$ supersymmetric $SU(5)$ model in five dimensions, where the
fifth dimension is compactified on $S^1/Z_2 \times Z'_2$. The circle
$S^1$ has coordinate $y$, with $y \equiv y + 2 \pi R$. Under the first $Z_2$, 
which maps $y \rightarrow 2 \pi R -y$, half of the supersymmetry charges 
are odd, so that $N=1$ supersymmetry is left in the four dimensional 
effective theory. The second $Z_2$, which we have denoted by a prime,
maps $y \rightarrow \pi R - y$. It is assumed that the gauge fields of 
$SU(3) \times SU(2) \times U(1)$ are even under $Z'_2$ and thus have 
massless Kaluza-Klein zero modes, whereas the remaining gauge fields --- 
those of $SU(5)/(SU(3) \times SU(2) \times U(1))$ --- are odd and thus have
only massive modes. Consequently, the low-energy 
effective theory has the symmetry of the supersymmetric standard
model. There are branes at $y=0$ (the fixed point of $Z_2$) and at
$y= -\pi R/2$ (the fixed point of $Z'_2$), which we will call
$O$ and $O'$ respectively. On the brane at $O$ there is a full $SU(5)$
local symmetry, whereas on the brane at $O'$ there is only a local
$SU(3) \times SU(2) \times U(1)$. 

We assume that on the brane at $O$ there are three families of quarks and 
leptons (i.e. three copies of ${\bf 10} + \overline{{\bf 5}}$),
which we will give a family index $i$. In the bulk, in addition to
the gauge fields of $SU(5)$, which are in a vector multiplet of $N=1$ 
5d supersymmetry, there are assumed to be 
two $\overline{{\bf 5}} + {\bf 5}$ pairs of hypermultiplets. 
One of these pairs of hypermultiplets, which we will denote 
$\overline{{\bf 5}}_H + 
{\bf 5}_H$, contains the two Higgs doublets of the MSSM. The other pair
of hypermultiplets, which we will denote $\overline{{\bf 5}}' + {\bf 5}'$,
contains extra vectorlike quark and lepton fields that mix with the three 
families living on the brane at $O$. In each of the hypermultiplets
$\overline{{\bf 5}}'$, ${\bf 5}'$, $\overline{{\bf 5}}_H$, and ${\bf 5}_H$,
there is a Kaluza-Klein tower of left-handed chiral 4d superfields that are
even under $Z_2$ (call them $\Phi$) and a tower of conjugate chiral 4d
superfields that are odd under $Z_2$ (call them $\Phi^c$). We assume that
in the $\Phi$ the weak doublets are even under $Z'_2$ and the 
color-triplets are odd. (The $Z'_2$ parities are necessarily opposite to 
this in $\Phi^c$.) Thus only the doublets in $\Phi$ are even under both $Z_2$ 
symmetries and have Kaluza-Klein zero modes. These doublet zero modes in 
$\overline{{\bf 5}}_H$ and ${\bf 5}_H$ are just the two Higgs doublets of the
MSSM. The doublet zero modes in $\overline{{\bf 5}}'$ and ${\bf 5}'$
are leptons and will get large masses by mixing with the fermions that
live on the brane at $O$. The low-energy spectrum, therefore, is
exactly that of the MSSM.

We see that this model is essentially a minimal $SU(5)$ model in five
dimensions, except for the presence of the fields 
$\overline{{\bf 5}}' + {\bf 5}'$. The bulk fields are assumed to have the
following couplings to the brane fields in the superpotential on the
brane at $O$: 
\begin{equation}
W_O  \supset  
\overline{\lambda}_{33} {\bf 10}_3 \overline{{\bf 5}}_3 
\overline{{\bf 5}}_H (O)  
+ \overline{M} \; \overline{{\bf 5}}'(O) {\bf 5}'(O)
+ \overline{m} \; \overline{{\bf 5}}_2 {\bf 5}'(O) + 
\overline{\lambda}_3 {\bf 10}_3 \overline{{\bf 5}}'(O) \overline{{\bf 5}}_H (O).
\end{equation}
\noindent
Note the similarity to Eq. (6)
The argument ($O$) after the bulk fields simply means that they are evaluated
at $y=0$ (the $O$ brane). The barred parameters are related to 
effective four-dimensional parameters (which have different mass dimension)
by $\overline{\lambda}_{33} =  \sqrt{\frac{\pi}{2} R} \lambda_{33}$, 
$\overline{M} = \frac{\pi}{2} R M$, $\overline{m} = \sqrt{\frac{\pi}{2} R} m$,
and $\overline{\lambda}_3 = \frac{\pi}{2} R \lambda_3$. The vacuum 
expectation value of the weak doublet Higgs is 
$\langle \overline{{\bf 5}}_H (O) \rangle = v_d/\sqrt{\frac{\pi}{2} R}$.
There are also terms, not shown, 
that give mass to the lighter families of quarks and leptons. 

In addition, there should be mass terms in the superpotential on $O'$
for those components of the fields $\overline{{\bf 5}}' + {\bf 5}'$ 
that do not vanish on that brane, i.e. for the doublets in the
$\Phi$ and the triplets in the $\Phi^c$:
\begin{equation}
W_{O'} \supset \overline{M}_2 \overline{{\bf 2}}'(O') {\bf 2}'(O') 
+ \overline{M}_3 \overline{{\bf 3}}^{\prime c} (O') 
{\bf 3}^{\prime c} (O'),
\end{equation}
\noindent
where, again, we define four-dimensional couplings by
$\overline{M}_2 = \frac{\pi}{2} R M_2$ and $\overline{M}_3 = 
\frac{\pi}{2} R M_3$.

\begin{figure}
\begin{center}
\includegraphics[width=4in]{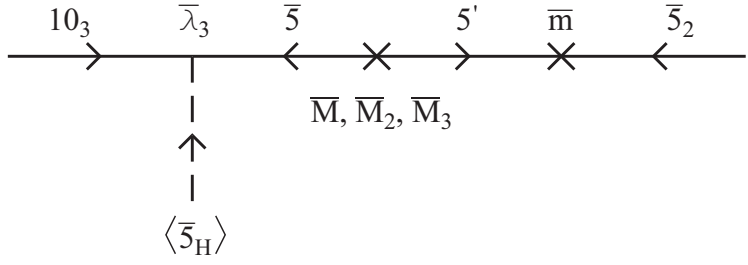}
\end{center}
\caption{\label{brane2} A diagram that yields the effective operators producing
``lopsided" contributions to $M_D$ and $M_L$ in five-dimensional theory. 
The fields $\overline{{\bf 5}}' + {\bf 5}'$
that are integrated out are the bulk fields and the cross in the 
middle of the diagram represents their allowed mass terms on both branes.}
\end{figure}

One may now proceed to integrate out the bulk quarks and leptons
in $\overline{{\bf 5}}' + {\bf 5}'$ as shown in Fig.~\ref{brane2}. If we take
into account only the Kaluza-Klein zero modes in these fields, then 
integrating them out gives a contribution to $M_L$ but not $M_D$, as only the
weak doublets in $\overline{{\bf 5}}' + {\bf 5}'$ have zero modes.
In that case one would find exactly the form of the matrices given
in Eq. (1). However, if we take into account also the infinite tower of
massive Kaluza-Klein modes, then both $M_L$ and $M_D$ get contributions
from Fig.~\ref{brane2}. The calculation can be done exactly and yields the following
result:
\begin{equation}
M_L \cong \left( \begin{array}{ccc} 0 & 0 & 0 \\ 0 & 0 & \rho \\
0 & 0 & 1 \end{array} \right) \; m, \;\;\;\; 
M_D \cong \left( \begin{array}{ccc} 0 & 0 & 0 \\ 0 & 0 & 0 \\
0 & \tilde{\rho} & 1 \end{array} \right) \; m,
\end{equation}
\noindent
where
\begin{equation}
\rho = (\lambda_3/\lambda_{33}) \frac{m}{\sqrt{m^2 [1 + \frac{1}{2}
(\frac{\pi}{2} R M_2)^2] + (M + M_2)^2}},
\end{equation}
\noindent
and 
\begin{equation}
\tilde{\rho} = \frac{1}{\sqrt{2}} (\lambda_3/\lambda_{33}) 
\frac{(\frac{\pi}{2} R m)(\frac{\pi}{2} R M_3)}
{\sqrt{[1 - (\frac{\pi}{2} R M)(\frac{\pi}{2} R M_3)]^2 + 
(\frac{\pi}{2} R m)^2 (\frac{\pi}{2} R M_3)^2 + (\frac{\pi}{2} R m)^2}}.
\end{equation}

The size of $\tilde{\rho}$ compared to $\rho$ depends on how large
the masses $m$, $M$, $M_2$ and $M_3$ that appear in the superpotentials
on $O$ and $O'$ are compared to the compactification scale $M_c = 1/R$.
And this, in turn, depends on where these masses are assumed to come from.
We assume that the five-dimensional model we are discussing is an effective
theory below some cutoff scale $M_s$. In that effective theory, one would 
expect that the dimensionless parameter $\overline{M} = \frac{\pi}{2} RM$ in 
Eq. (8) would be of order 1. However, if $\overline{M}$ arises from the 
GUT-scale (i.e. $O(M_c)$) vacuum expectation value of some bulk singlet 
field ${\bf 1}_H$ through a term 
$\overline{\lambda} {\bf 1}_H (O) \overline{{\bf 5}}' (O) {\bf 5}' (O)$,
then one would expect $\overline{\lambda} \sim M_s^{-3/2}$ and thus
$\frac{\pi}{2} R M$ to be of order $(M_c/M_s)^{3/2}$. (If $\overline{M}$
arises from the GUT-scale VEV of a singlet that lives on the brane at
$O$, then one would expect $\frac{\pi}{2} R M \sim M_c/M_s$.)
In the same way, if the other masses, $m$, $M_2$, and $M_3$, come from
singlet VEVs, they would also naturally be small compared to $M_c$.
If we do assume that these masses are suppressed in this way, then one 
obtains the result that $\tilde{\rho} \ll \rho \sim 1$, which realizes 
the structure given in Eq. (1). 

In a five-dimensional model, such as this, there are corrections to
$m_{\tau}/m_b$ coming from five-dimensional effects that have to be considered.
Let us consider the form of the matrices $M_D$ and $M_L$ at the 
compactification scale $M_c = 1/R$. At that scale the 33 elements of
these matrices should be equal, since the RGE corrections to these
parameters above $M_c$ should respect $SU(5)$. (Above $M_c$, one may
use a five-dimensional description of the theory. In that description, 
the bulk terms and the terms on the brane at $O$ respect $SU(5)$. There
can be terms on the brane at $O'$ that explicitly violate $SU(5)$, since
there is no local $SU(5)$ at $O'$. However, these have no effect on the
33 elements of $M_D$ and $M_L$, as these arise from Yukawa terms on the
brane at $O$.) Thus, at the scale $M_c$, there is the relation 
$m_{\tau}/m_b = \sqrt{1 + \tan^2 \theta^{\ell}_{23}}$
that we discussed earlier. Below $M_c$, the ratio $m_{\tau}/m_b$ will 
run down to low
energies as in the four-dimensional minimal $SU(5)$ theory. However,
it only runs between $M_c$ and the weak scale \cite{Hall:2002ci}, whereas in a 
four-dimensional theory it runs between $M_{GUT}$ and the weak scale.
Since the value of $M_{GUT}$ in the four-dimensional minimal $SU(5)$ theory
is not the same as the value of $M_c$ in the five-dimensional 
$SU(5)$ theory, there are corrections to $m_{\tau}/m_b$ relative to
the four-dimensional theory that are of order $\alpha_3/\pi \ln (M_{GUT}/M_c)$.
These are largest if $M_c$ is small compared to the fundamental scale
$M_s$ and thus compared to $M_{GUT}$ (i.e. for widely separated branes).
Hall and Nomura \cite{Hall:2002ci} compute these five-dimensional corrections to 
$m_{\tau}/m_b$ to be $- 4 \%$ for the extreme case where $M_s/M_c \sim 
(4\pi)^2$. This is not enough to restore agreement with the minimal
$SU(5)$ prediction.

The structure shown in Eq. (10), and the extension of it with large
$(M_L)_{13}$, can be embedded in a model
of the quark and lepton masses of all three families, as we will now show 
by a simple example. Suppose that there is a $U(1)$ flavor symmetry, broken 
spontaneously by the vacuum expectation value of a flavon field, $\phi_f$, 
whose flavor charge is $-1$. Entries in the quark and lepton mass matrices
that violate the $U(1)$ charge by $n$ units will thus ``cost" $n$ powers
of the flavon field, and thus be proportional to a small symmetry-breaking 
parameter $\epsilon^n$. This is just the usual Froggatt-Nielson kind 
of scenario. Assign the $U(1)$ charges as follows:
${\bf 10}_1(+4)$, ${\bf 10}_2(+2)$, ${\bf 10}_3(0)$, 
$\overline{{\bf 5}}_1(+4)$, $\overline{{\bf 5}}_2(+4)$,
$\overline{{\bf 5}}_3(+3)$, $\overline{{\bf 5}}'(0)$, 
${\bf 5}'(-4)$, $\overline{{\bf 5}}_H(0)$, and
${\bf 5}_H(0)$. Let there also be a flavon field $\phi'_f$ localized on the 
brane at $O'$ having $U(1)$ flavor charge $+4$. This last field allows
the mass terms for the fields $\overline{{\bf 5}}'$ and ${\bf 5}'$
that are given in Eq. (9). All the terms in Eq. (8) are allowed 
(with suitable powers of the flavon field $\phi_f$ and thus of $\epsilon$)
except the second term. Thus the mass matrices have the form
\begin{equation}
\begin{array}{ll}
M_L = \left( \begin{array}{ccc} a_{11} \epsilon^8 & a_{12} \epsilon^6 & 
\rho' + a_{13} \epsilon^4 \\ 
a_{21} \epsilon^8 & a_{22} \epsilon^6 & \rho + a_{23} \epsilon^4 \\
a_{31} \epsilon^7 & a_{32} \epsilon^5 & a_{33} \epsilon^3 \end{array} 
\right) m_D, & 
M_D = \left( \begin{array}{ccc} a_{11} \epsilon^8 & a_{21} \epsilon^8 & 
a_{31} \epsilon^7 \\ a_{12} \epsilon^6 & a_{22} \epsilon^6 & a_{32} \epsilon^5
\\ \tilde{\rho}' + a_{13} \epsilon^4 & \tilde{\rho} + a_{23} \epsilon^4 & 
a_{33} \epsilon^3 \end{array} \right) m_D, \\  & \\
M_U = \left( \begin{array}{ccc} b_{11} \epsilon^8 & b_{12} \epsilon^6 &
b_{13} \epsilon^4 \\ b_{12} \epsilon^6 & b_{22} \epsilon^4 & b_{23} \epsilon^2
\\ b_{13} \epsilon^4 & b_{23} \epsilon^2 & b_{33} \end{array} \right) m_U, &
M_{\nu} = \left( \begin{array}{ccc} c_{11} \epsilon^8 & c_{12} \epsilon^8
& c_{13} \epsilon^7 \\ c_{12} \epsilon^8 & c_{22} \epsilon^8 &
c_{23} \epsilon^7 \\ c_{13} \epsilon^7 & c_{23} \epsilon^7 & c_{33} \epsilon^6
\end{array} \right) m_U^2/m_R \\ & \end{array}
\end{equation}
\noindent
If we define $\sigma \sim \sqrt{\rho^2 + \rho^{\prime 2}} \sim \epsilon^3$, so that
the atmospheric neutrino mixing angle is of order one,
then we see that $m_{\mu} \sim \epsilon^5$
and $m_s \sim \epsilon^6$.
So this kind of model, being much more lopsided in the charged lepton
sector than in the down quark sector, naturally accounts for the fact
that $m_{\mu}$ is much larger than $m_s$ at the GUT scale. The Georgi-Jarlskog
factor \cite{gj}
($m_s/m_{\mu}$) comes out naturally of order $\epsilon \sim 0.2$,
though in this model it is not precisely predicted. 

Note that in this particular model there are contributions to 
$\theta_{atm}$ that are of order $\epsilon$ coming from the 
diagonalization of $M_{\nu}$. This
means that no precise and testable relation of the kind we want exists between
$\theta_{atm}$ and $m_{\tau}/m_b$, unless the model is augmented in some
way as to make 
the $O(1)$ coefficients $c_{33}$, $c_{23}$, etc. in Eq. (13)
predictable or small.

\begin{table}
\caption{\label{tab:table1}The value of $m_{\tau}/m_b$ at the GUT scale versus
$\tan \beta$ in the MSSM. The values $m_b(m_b) = 4.25$ GeV and 
$m_{\tau}(m_{\tau}) = 1.777$ GeV are used. 3-loop QCD and 1-loop QED RGE 
equations are used in running up to $m_t$. All SUSY particle masses are 
taken to be degenerate at $m_t$. From $m_t$ to $M_{GUT}=2 \times 10^{16}$ GeV
the 2-loop MSSM beta functions are used. The gauge couplings are taken to
be $\alpha^{-1} (M_Z) = 127.9$, $\alpha_2^{-1} (M_Z) = 29.61$, and 
$\alpha_3 (M_Z) = 0.118$.}
\begin{ruledtabular}
\begin{tabular}{c@{\hspace{.5cm}}c@{\hspace{1.5cm}}c@{\hspace{.5cm}}c@{\hspace{1.5cm}}c@{\hspace{.5cm}}c}
$\tan \beta$&$m_{\tau}/m_b$ &$\tan \beta$&$m_{\tau}/m_b$& $\tan \beta$& $m_{\tau}/m_b$ \\
\hline
1.85 & 1.03 & 5.0  & 1.20 & 50.0 & 1.11 \\
2.0  & 1.09 & 10.0 & 1.21 & 55.0 & 1.06\\
2.5  & 1.15 & 20.0 & 1.20 & 58.0 & 1.0\\
3.0  & 1.17 & 40.0 & 1.16 \\
\end{tabular}
\end{ruledtabular}
\end{table}

\end{document}